\documentclass[prc,aps,preprint,showpacs]{revtex4}
\usepackage{graphicx}
\newcommand{\simgeq}{\; \raisebox{-0.4ex}{\tiny$\stackrel
{{\textstyle>}}{\sim}$}\;}

\begin{document}
\title{Neutron shell structure and deformation in neutron-drip-line nuclei} 

\author{ Ikuko Hamamoto$^{1,2}$ }

\affiliation{
$^{1}$ {\it Riken Nishina Center, Wako, Saitama 351-0198, Japan } \\ 
$^{2}$ {\it Division of Mathematical Physics, Lund Institute of Technology 
at the University of Lund, Lund, Sweden} }   




\begin{abstract}
Neutron shell-structure and 
the resulting possible deformation in the neighborhood of  
neutron-drip-line nuclei are systematically discussed, 
based on both bound and resonant neutron one-particle energies 
obtained from spherical and deformed Woods-Saxon potentials. 
Due to the unique behavior of weakly-bound and resonant neutron one-particle
levels with smaller orbital angular-momenta $\ell$, a systematic change of the
shell structure and thereby the change of neutron magic-numbers are pointed out, 
compared with those of stable nuclei expected from the conventional j-j 
shell-model. 
For spherical shape with the operator of the 
spin-orbit potential conventionally used, 
the $\ell_{j}$ levels belonging to a given oscillator major shell 
with parallel spin- and orbital-angular-momenta 
tend to gather together in the energetically 
lower half of the major shell, 
while those levels with anti-parallel spin- and
orbital-angular-momenta gather in the upper half. 
The tendency leads to a unique shell structure and 
possible deformation when neutrons start to occupy the orbits 
in the lower half of the major shell. 
Among others, the neutron magic-number N=28 disappears and N=50  
may disappear, while the magic number N=82 may presumably survive due to
the large $\ell =5$ spin-orbit splitting for the $1h_{11/2}$ orbit.    
On the other hand, an appreciable amount of energy gap may appear 
at N=16 and 40 for spherical shape, 
while neutron-drip-line nuclei 
in the region of neutron number above N=20, 40 and 82, namely
N $\approx$ 21-28, N $\approx$ 41-54, and
N $\approx$ 83-90, may be quadrupole-deformed though the possible deformation
depends also on the proton number of respective nuclei.    
\end{abstract}


\maketitle

\newpage

\section{INTRODUCTION} 
Thanks to the development of various facilities of radioactive nuclear ion beam,
the knowledge of nuclei far away from the stability line has recently been   
much increased.  Though the neutron drip line has so far 
been experimentally pinned down up till the oxygen (Z=8) isotope,
at the moment the experimental knowledge of nuclei with Z$>$8 
close to the neutron drip line is quickly increasing.

The study of unstable nuclei, especially neutron-drip-line nuclei which contain 
very weakly bound neutrons, has opened a new field in the study of the
structure of finite quantum-mechanical systems.  The study is important 
not only for the interests in nuclear astrophysics namely for the 
understanding of the production of energy and the
synthesis of elements in stars and during stellar events, but also for giving
the opportunity for learning the properties of the Fermion system with  
very loosely-bound particles, some density of which can extend to the region 
far outside the region of the main density of the system. 
Various exciting study of man-made finite quantum-mechanical 
systems, such as clusters of atoms and quantum dots, has recently been   
made possible, however, those systems are so far limited to be well 
bound so that the related potentials are in a good approximation 
simulated by a harmonic-oscillator.  
Since the nucleon separation energy in stable nuclei is 7-10 MeV, the
spectroscopic analysis around the ground state of stable nuclei 
has been successfully carried out  
also in terms of harmonic-oscillator wave-functions. Correspondingly, most of
systematic nuclear shell-model calculations are so far carried out using 
harmonic-oscillator wave-functions.

The study of one-particle motion in deformed potentials is the basis for
understanding the structure of deformed nuclei.  Since the Fermi level
of drip line nuclei lies close to the continuum, both weakly-bound and
positive-energy one-particle levels play a crucial role in the many-body
correlations of those nuclei.  Among an infinite number of one-particle levels
at a given positive energy, only some selected levels related to one-particle
resonant levels will be important for the understanding of the properties of 
bound states of drip-line nuclei.  

The behavior of $s$ neutrons is an extreme example because 
the barrier coming from either centrifugal or Coulomb potentials is absent. 
Therefore, for example, 
as the separation energy approaches zero, the probability of 
$s$ neutrons staying inside the nuclei approaches zero.       
When a larger part of a bound one-particle wave-function lies 
outside the nuclear 
potential, the one-particle eigen energy becomes less sensitive to the  
potential provided by the well-bound nucleons 
in the system.
In contrast, the wave functions of weakly-bound but large-$\ell$ neutrons 
stay mostly inside the nuclear potential, due to the high barrier coming from
the centrifugal potential, of which the height is proportional to $\ell
(\ell +1)$. 
Consequently, when the potential changes (or the neutron number for a given
proton number approaches the neutron-drip-line) so that 
one-particle energies of last bound neutrons approach zero, 
the binding energy of larger-$\ell$ neutrons, which is more sensitive to the
strength of the potential, decreases much faster than that 
of smaller-$\ell$ neutrons.  
The height of the centrifugal potential also affects sharply 
the properties of one-particle resonant levels.   
Thus, the same behavior of $\ell$-dependence as that of weakly-bound
neutron energies is  
obtained also for lower-lying neutron one-particle 
resonant energies, as shown in Ref. \cite{IH04, IH05, IH06}.
This $\ell$-dependent behavior of one-particle energies on the potential
strength  
leads to a systematic change of the
shell structure in both weakly-bound and resonant neutron 
one-particle energies compared with the shell structure 
of strongly-bound neutrons
\cite{IH07}.  

Using the numerical result of the self-consistent mean-field calculations 
with effective interactions used in stable nuclei 
while limiting the spherical system in a
large box, one-particle spectra of bound nucleons in spherical 
drip-line-nuclei were
studied in Ref. \cite{JD94}.  
And, some systematic change of shell structure in bound neutron energies 
of spherical neutron drip-line nuclei was obtained, which is similar
to that of the present work found at $\beta = 0$   
except the conclusion in Ref. \cite{JD94} that the presence of
magic gaps at the neutron number 
N = 28, 50 and 82 does not appreciably change as one approaches the
neutron-drip-line.

Taking the $(N=2)$ $sd$- and $(N=3)$ $pf$-shells where $N$ expresses the
harmonic-oscillator principle quantum-number, 
the systematic change of shell structure due to the unique behavior of  
one-particle energies of weakly-bound or resonant
levels with small $\ell$ is briefly explained in the following.  
In stable $sd$-shell nuclei, it is well-known that  
the relation of one-particle energies is such that 
$\epsilon(1d_{5/2}) < \epsilon(2s_{1/2}) < \epsilon(1d_{3/2})$, which 
agrees with experimental information.  
However, in lighter neutron-rich nuclei, in which the $2s_{1/2}$ level
becomes very weakly bound, the relation  
$\epsilon(2s_{1/2}) \approx \epsilon(1d_{5/2}) < \epsilon(1d_{3/2})$ is expected, 
which leads to the new magic number N=16 \cite{AO00}.  In fact, the appearance
of the N = 16 neutron magic number for unstable nuclei together with a
weakening of the shell closure at N = 20 and 28 was 
mentioned already in 1975 \cite{BL75} based on the self-consistent calculations 
using the energy density formalism with pairing interaction. 
On the other hand, in heavier nuclei where neutrons
in the $sd$-shell are strongly bound, the relation 
$\epsilon(1d_{5/2}) < \epsilon(1d_{3/2}) < \epsilon(2s_{1/2})$
is obtained from both Hartree-Fock calculations and eigenvalues of 
Woods-Saxon potentials with a practical strength of spin-orbit potential.  
Similarly, if we take an example of the $pf$-shell, the relation  
$\epsilon(1f_{7/2}) < \epsilon(1f_{5/2}) < \epsilon(2p_{3/2}) 
< \epsilon(2p_{1/2})$ is obtained in the case of strongly-bound $pf$-neutrons. 
For stable $pf$-shell nuclei, the relation  
$\epsilon(1f_{7/2}) < \epsilon(2p_{3/2}) < \epsilon(1f_{5/2}) \approx 
\epsilon(2p_{1/2})$ is known.  However, when one-particle levels of 
$1f_{7/2}$ and $2p_{3/2}$ become very weakly-bound or resonant, the relation  
$\epsilon(1f_{7/2}) \approx \epsilon(2p_{3/2})$ appears.  The relation leads to
the disappearance of the magic number N = 28 \cite{IH07} 
and, moreover, nuclei with some
neutrons in the almost degenerate $f_{7/2}$ and $p_{3/2}$ shells, which couple
strongly each other by quadrupole-quadrupole interaction, may be easily
quadrupole deformed.  The degeneracy can well be responsible for the presence  
of ''island of inversion''.   It has been known that the presence of only like
nucleons in a large single-j-shell can hardly lead to quadrupole 
deformation.  
On the other hand, the presence of like nucleons in 
several nearly degenerate j-shells,   
which couple strongly each other by quadrupole-quadrupole interaction, 
may lead to quadrupole deformation.  

In axially-symmetric quadrupole-deformed nuclei 
the role of smaller-$\ell$ neutrons for spherical shape is replaced by neutrons
with smaller $\Omega$ values, where $\Omega$ denotes the angular momentum
component of neutrons along the axially-symmetry axis.  
For example, the smallest
possible angular-momentum component of $\Omega^{\pi} = 1/2^{+}$ orbits 
is $s_{1/2}$, which becomes always the 
overwhelming component of 
angular-momentum in neutron one-particle wave-functions 
with $\Omega^{\pi}$ = 1/2$^+$  as the binding energy 
of the neutron approaches zero \cite{MT97, IH04}.  
In the case that the smallest orbital angular-momentum is
not equal to zero, it depends on the properties of respective orbits how the
component of the smallest orbital angular-momentum becomes dominant 
in one-particle wave-functions when the binding energy approaches zero
\cite{IH04}. 
Since all spherical one-particle orbits with positive parity 
($s_{1/2}, d_{3/2}, d_{5/2}, ...$) have 
an $\Omega^{\pi} = 1/2^{+}$ component, the shell structure change for deformed
shape close to the continuum due to the unique property of $s_{1/2}$ orbit
is expected to
occur more often compared with the case of spherical shape. 
It is the most convenient way to see the shell structure of deformed nuclei 
to plot one-particle energies as a function of quadrupole deformation 
(Nilsson diagram) \cite{BM75}.  
Therefore, in this article Nilsson diagrams that are relevant to some
possible neutron-drip-line nuclei related to neutron magic numbers 
in stable nuclei are presented.
The change of nuclear shell structure for neutrons is seen in both bound and
resonant one-particle energies in Nilsson diagrams.

In Sec. II main points of our model are briefly summarized, while numerical
results are presented in Sec. III.  Conclusions and discussions are given in
Sec. IV.

\section{MODEL}
In order to solve the eigenvalue \cite{IH04} and eigenphase \cite{IH05,IH06} 
problems for neutron one-particle bound and resonant levels, respectively, 
as a function of axially-symmetric quadrupole deformation, 
the coupled differential equations obtained from the 
Schr\"{o}dinger equation are integrated in coordinate space with correct 
asymptotic behavior at $r = R_{max}$ , where $R_{max}$ is so large that 
both the
nuclear potential and the coupling term are negligible.  
In this way one-particle resonant energy in deformed nuclei  
can also be estimated without any ambiguity.   
For $\beta \neq 0$ the resonant energy is defined as the energy, at
which one of the eigenphases increases through $\pi$/2 as the energy increases 
\cite{RGN66, IH05, IH06}.  
One-particle resonance is absent if none of eigenphases
increase through $\pi$/2 as the energy increases. For example, neutron  
one-particle resonant state with $\Omega^{\pi}$ = 1/2$^{+}$ is not obtained 
as far as  the major component of the one-particle wave-function comes from 
$\ell = 0$, because an $\Omega^{\pi}$ = 1/2$^{+}$ level with an appreciable
amount of the $\ell$=0 component can very quickly decay via the $\ell$=0 channel. 
Since the height of the centrifugal barrier decreases for a larger nuclear
radius, the unique behavior of $\ell$=1 components contained in the 
$\Omega^{\pi}$ = 1/2$^{-}$ and 3/2$^{-}$ orbits will be more easily seen 
in heavier nuclei. 

On the other hand, since the height of the centrifugal barrier is 
proportional to $\ell (\ell +1)$, at a given positive energy for a given 
potential  
the width of a one-particle resonant level is larger for the level with smaller
orbital angular-momentum.  As the energy increases the width of a given
resonant level becomes larger, and finally at a certain energy 
the one-particle level with a given $\ell$ is 
no longer obtained as a resonant level.  
For simplicity, 
the calculated widths of one-particle resonant levels are not given in the
present article, 
since the widths are not a major subject in this work.

Though weakly-bound neutrons in nuclei close to the 
neutron drip line have a contribution especially 
to the tail of the self-consistent nuclear potentials, 
the major part of the
nuclear potential is provided by well-bound nucleons, 
especially by strongly-bound protons in the case of neutron-rich nuclei.  
Therefore, for simplicity, in this article 
the parameters 
of Woods-Saxon potentials are taken from the standard ones (see p.239 of 
\cite{BM69}).  Namely, 
the diffuseness $a$=0.67 fm, the radius $r_0 A^{1/3}$ where $r_0$=1.27 fm,
the depth of the Woods-Saxon potential for neutrons is  
\begin{equation}
V = -51 + 33 \, \frac{N-Z}{A}  \qquad \mbox{MeV}
\end{equation}
and the spin-orbit potential is expressed by 
\begin{equation}
V_{\ell s} = -0.44 V \, (\vec{\ell} \cdot \vec{s}) \, r_0^{2} \, 
\frac{1}{r} \, \frac{d}{dr} f(r) \qquad \mbox{MeV}
\end{equation}
where 
\begin{equation}
f(r) = \frac{1}{1 + exp(\frac{r-R}{a})}
\end{equation}
It is noted that the neutron potential for nuclei with a neutron excess is 
shallower than that for N=Z nuclei.
In fact, the nuclear potential with the above set of parameters 
is found to approximately reproduce the position of the neutron drip line, 
which is expected from presently available experimental data.

In the discussion of possible deformation of given nuclei examining the
Nilsson diagram  
we use the following empirical facts: 
(a) If pair correlation plays a minor role,
the presence of neutrons in almost degenerate $\ell_{j}$ shells 
around the Fermi level may make the system deformed 
since those neutrons have a possibility of gaining energy by breaking spherical
symmetry (Jahn-Teller effect);  
(b) In order to obtain a deviation from spherical shape, the energies of 
one-particle levels  
just below and on the Fermi level in the Nilsson diagram 
need to be mostly decreasing (downward-going) 
for $\beta = 0 \rightarrow \beta \neq 0$ so that the system gains the energy by
deformation \cite{BM75};
(c) The presence of only like nucleons in a large single-j-shell 
may not be sufficient to deform the system, though the presence of
both neutrons and protons in a given single-j-shell may induce some quadrupole
deformation.  Examples are: no observed deformed nuclei both in the $_{38}$Sr 
and $_{40}$Zr isotopes due to the occupation of the $1g_{9/2}$ shell 
by neutrons \cite{GJK12},
and in the $_{18}Ar$ and $_{20}$Ca isotopes due to the occupation of the
$1f_{7/2}$ shell by neutrons; 
(d) Only prolate deformation is discussed,
since it is empirically known that prolate deformation is overwhelmingly 
dominant among deformed nuclei though the absolute dominance is not yet fully
understood \cite{HM09}.

\section{NUMERICAL RESULTS}
Though the near degeneracy of both 
the $1d_{5/2}-2s_{1/2}$ levels in the $N$=2 oscillator shell and 
that of $1f_{7/2}-2p_{3/2}$ levels in the $N$=3 shell of 
the spherical potential in neutron-drip-line nuclei as well as the resulting
possible deformation are partly discussed in Ref. \cite{IH07}, in the
following we include a brief description of those cases for completeness.  
The two remaining $n \ell_{j}$ levels, 
$1f_{5/2}$ and $2p_{1/2}$, in the $N$ = 3 oscillator major shell
other than the $1f_{7/2}$ and $2p_{3/2}$ 
levels, in which spin and orbital angular
momenta are anti-parallel, may become also almost degenerate around the Fermi
level of certain nuclei. Nevertheless,
the degeneracy may not lead to a deformation, 
because those levels lie in the
second half of the $N$ = 3 major shell and, then, 
the general behavior of the deformed one-particle energies originating
from those levels in the spherical limit 
is energetically upward-going for  
$\beta = 0 \rightarrow \beta \neq 0$. 
The relation between deformation of the system and upward-going (or
downward-going) energy levels 
in the Nilsson diagram is known already in the study of stable rare-earth 
nuclei \cite{BM75}.  Namely, 
energetically downward-going one-particle levels  
for $\beta = 0 \rightarrow \beta > 0$ (prolate shape) 
around N $\simgeq$ 88-90 that lead to stable deformed rare-earth nuclei, 
end at  N $\approx$ 110, around
which 
the observed deformation of stable rare-earth nuclei also ends.   
(See, for example, Fig.5-3 of \cite{BM75}.)
As is seen in the following, the shell structure unique in neutron 
weakly-bound and resonant levels leads to the bunching of one-particle levels in
a given $N$ major shell for spherical shape: 
levels with parallel spin- and orbital-angular-momenta 
gather together in the lower half of
the major shell, 
while levels with anti-parallel spin- and orbital-angular-momenta 
gather in the upper half shell. Levels within the respective groups couple 
each other strongly by spin-independent quadrupole-quadrupole interaction.  
However, we note that there is 
a difference between the two groups concerning the
possible deformation.  Namely, having neutrons in the nearly degenerate levels 
with parallel spin- and orbital- 
angular-momenta may make the system deformed, because one-particle
energies in the lower half of a given $N$ major shell are in general 
decreasing for $\beta = 0 \rightarrow \beta \neq 0$, as can be seen from
the Nilsson diagram.  
Since the levels belonging to each group have different orbital 
angular-momenta, the occurrence of near degeneracy depends on the actual 
strength of spin-orbit splitting and the values of relevant orbital angular
momenta.  As seen in the $1h_{11/2}$ orbit of Fig. 4, the highest-j level with
parallel spin- and orbital-angular-momenta  
tends to go out of the group of degenerate levels in heavier nuclei.

\subsection{Near degeneracy of $1d_{5/2}$ and $2s_{1/2}$ levels}
In Fig. 1 we show the Nilsson diagram for neutrons, in which parameters of 
the Woods-Saxon potential are chosen for the nucleus $^{18}_{6}$C$_{12}$.
It is noted that the observed ground-state spins of nuclei $^{15}_{6}$C$_{9}$, 
$^{17}_{6}$C$_{11}$, and $^{19}_{6}$C$_{13}$ are 1/2$^+$, 3/2$^+$, and 1/2$^+$, 
respectively, and are most easily understood in terms of prolate deformation  
for $\beta >$ 0.1, where the last odd neutron occupies 
the $\Omega^{\pi}$ = 1/2$^+$, 3/2$^{+}$ and 
1/2$^{+}$ levels that correspond to  
the 9th, 11th, and 13th neutron one-particle levels, respectively, 
assuming that the respective even-even core nucleons occupy pair-wise the 
lower-lying Nilsson one-particle levels and 
couple to $I^{\pi}$ = $K^{\pi}$ = 0$^+$.   
For some experimental evidence for the deformation of those C-isotopes, see
Refs. \cite{ZE04,ZE05}.                
At $\beta$ = 0 the calculated energy difference 
between the $2s_{1/2}$ and $1d_{5/2}$
levels in Fig. 1 is only 509 keV.  In contrast, 
a large energy gap for spherical shape ($\beta$ = 0) 
appears at N = 16, since the calculated $1d_{3/2}$ resonant energy is 4.36 MeV.

\subsection{Near degeneracy of $1f_{7/2}$ and $2p_{3/2}$ levels} 
In Fig. 2 the Nilsson diagram for neutrons is shown, in which parameters of the 
Woods-Saxon potential are chosen for the nucleus $^{34}_{12}$Mg$_{22}$.
At $\beta$ = 0 the calculated energy difference between the very weakly 
bound $1f_{7/2}$ level and very low-lying one-particle resonant 
$2p_{3/2}$ level is only 387 keV, 
which clearly indicates that the N = 28 energy gap at 
$\beta$=0 disappears in neutron-drip-line nuclei.  
This near degeneracy of the two levels, which couple strongly each other 
by spin-independent quadrupole-quadrupole interaction, may lead to the
deformation of the system with N $\approx$ 21-28, 
as a result of the Jahn-Teller effect.  In particular, odd-A nuclei
with N = 21 such as $^{33}_{12}$Mg$_{21}$ \cite{DTY07, DTY10} 
and $^{31}_{10}$Ne$_{21}$ \cite{TN09}
are observed to be deformed, being consistent with the strongly down-sloping
Nilsson one-particle levels with $\Omega^{\pi}$ = 1/2$^-$ 
and 3/2$^-$ for $\beta > 0$, 
which originate from
the $1f_{7/2}$ shell in the spherical limit ($\beta$ = 0).  
Neutron-drip-line nuclei with N = 21-28 
can well be deformed though the possible deformation depends also on the proton
number of respective nuclei.  
Examining the Nilsson diagram for protons it is seen that the proton numbers Z =
9 (F), 10 (Ne), 11(Na) and 12 (Mg) may prefer some deformation 
since the energies of the two lowest-lying
Nilsson one-proton levels in the $sd$-shell decrease sharply as 
$\beta = 0 \rightarrow \beta \neq 0$.    
The recent 
experimental information on the Mg isotope with N = 21-26 \cite{PD12} seems 
to go well with this interpretation using the Nilsson diagram. 
Indeed, this near degeneracy of the $1f_{7/2}$ and $2p_{3/2}$ levels
can be an important element for creating ''island of inversion''.  
In other words, heavier nuclei in the ''island of inversion'' could survive
inside the neutron-drip-line thanks to the deformation.

From Fig.2 it is also seen 
that at $\beta$=0 the well-bound $2s_{1/2}$ level lies
approximately in the middle of the $2d_{5/2}$ and $2d_{3/2}$ levels, in contrast
to the $sd$-shell level scheme shown in Fig.1.

\subsection{Near degeneracy of $1g_{9/2}$, $3s_{1/2}$ and $2d_{5/2}$ levels}
In Fig. 3 the Nilsson diagram for neutrons is shown, in which parameters of the
Woods-Saxon potential are chosen for the nucleus $^{66}_{22}$Ti$_{44}$.
A considerable amount of energy gap appears at N = 40 for spherical shape,
while the possible location of the $3s_{1/2}$ level slightly above zero
(indicated by the open circle in Fig. 3) is obtained from  
the extrapolation of 
the bound one-particle energy level 
with $\Omega^{\pi}$ = 1/2$^+$ for $\beta > 0.12$
denoted by the solid curve in Fig. 3, which reaches zero at $\beta$ 
= 0.12, since the continuation of the one-particle resonant level 
to the region of $\beta < 0.12$ cannot be obtained
due to the predominant $\ell$ = 0 component of the orbit.  Thus, the calculated 
$1g_{9/2}$, $3s_{1/2}$ and $2d_{5/2}$ levels, 
which are the three $n \ell_{j}$ levels belonging to the
$N$ = 4 oscillator shell with parallel spin- and orbital-angular-momenta and 
couple strongly each other by
spin-independent quadrupole-quadrupole interaction, lie within 1.43 MeV.  
This means that the energy gap at the magic number N = 50 clearly disappears.

The strong quadrupole coupling of these three levels can be seen 
from the Nilsson
diagram in Fig. 3.  For example,  
the lowest-lying one-particle level with 
$\Omega^{\pi}$ = 1/2$^{+}$ for $\beta > 0$, 
which is connected to the $1g_{9/2}$ shell at
$\beta$ = 0, contains a considerable amount of $3s_{1/2}$ and 
$2d_{5/2}$ components
already at moderate values of $\beta$.   
This 
can be seen from the comparison between the slopes of the $\Omega^{\pi}$ =
1/2$^{+}$ curve for $\beta > 0$ and the $\Omega^{\pi}$ = 9/2$^+$ curve 
for $\beta < 0$, both of which originate 
from the $1g_{9/2}$ level at $\beta$ = 0.  
The wave-function of the 
$\Omega^{\pi}$ = 9/2$^+$ orbit is almost pure $1g_{9/2}$ 
in the present range of
$\beta$-values since there is
no $\Omega^{\pi}$ = 9/2$^+$ one-particle orbit in the neighborhood. 
If both orbits, $\Omega^{\pi}$ = 1/2$^+$ for $\beta > 0$ and 
$\Omega^{\pi}$ = 9/2$^+$ for $\beta < 0$, 
consist only of a single-j-shell, namely $g_{9/2}$, then, 
the absolute magnitude 
of the slope, $| \frac{d \epsilon_{\Omega}}{d \beta} |$, of the $\Omega^{\pi}$ = 
9/2$^+$ level is a factor 2 larger than that of the $\Omega^{\pi}$ = 1/2$^+$
level. 
On the other hand, for a pure harmonic oscillator potential (namely the strong
mixing limit) the former is a half (= 0.5) of the latter.   
In Fig. 3 the ratio of the former to the latter is about 2, of course, 
for $|\beta| << 1$, 
while the absolute magnitude of the slope of the
$\Omega^{\pi}$ = 1/2$^+$ energy level becomes larger than that of the 
$\Omega^{\pi}$ = 9/2$^+$ energy level already at $|\beta|$ smaller than 0.3.  

The near degeneracy of these three levels, $1g_{9/2}$, $3s_{1/2}$ and
$2d_{5/2}$, corresponds to that of 
the $1f_{7/2}$ and $2p_{3/2}$ levels discussed in the previous subsection, which
leads to ''the island of inversion''.
First of all, neutron-drip-line nuclei with N = 41 is likely to be deformed, 
though the possible deformation depends also on the proton number of respective
nuclei.  
The ground-state spin of 
neutron-drip-line nuclei with N = 41 can be either 1/2$^{+}$ or 5/2$^{+}$ or 
1/2$^{-}$ or 3/2$^{-}$ depending on $\beta$-values 
if they are prolately deformed, instead of 9/2$^{+}$
expected for spherical shape.  
Secondly, a system having several neutrons in these  
($1g_{9/2}$ - $3s_{1/2}$ - $2d_{5/2}$) almost degenerate shells such as 
N $\approx 41-54$ is likely to be
deformed in a similar way to ''the island of inversion'', when pairing
interaction plays a minor role.  
See the fourth paragraph of Sec. IV for the discussion of the role of pairing
interaction in the determination of the shape of neutron-drip-line nuclei. 

The near degeneracy of the $1g_{9/2}$ level with other two levels has occurred 
for the phenomenological strength of the spin-orbit splitting which is not 
yet so strong for the $1g_{9/2}$ orbit. As is seen in the next subsection, 
in the $N$ = 5 oscillator major shell 
the spin-orbit splitting of the $1h_{11/2}$ level is 
so strong that the similar degeneracy of all levels with parallel 
spin- and orbital-angular-momenta belonging to the $N$ = 5 major shell 
can hardly occur.   

The remaining two levels in the $N$ = 4 oscillator major shell, 
$1g_{7/2}$ and $2d_{3/2}$, may be almost degenerate around the Fermi level
of certain neutron-drip-line nuclei.  However, it may be difficult to gain
energies by deforming those nuclei which have neutrons 
in those two levels, since one-particle
energies originating from the $1g_{7/2}$ and $2d_{3/2}$ levels 
are located in the second half 
of the $N$ = 4 major shell and thus the energies of the majority of one-particle
levels increase 
for $\beta = 0 \rightarrow \beta \neq 0$.

\subsection{Near degeneracy of $1h_{11/2}$, $2f_{7/2}$ and $3p_{3/2}$ levels ?} 
In Fig. 4 the Nilsson diagram for neutrons is shown, in which parameters of the
Woods-Saxon potential are chosen for the nucleus $^{126}_{44}$Ru$_{82}$.
A considerable amount of the energy gap remains at N = 82 for spherical shape
due to the large spin-orbit splitting of the $\ell$ = 5 level, $1h_{11/2}$, 
while the
very weakly-bound $2f_{7/2}$ and $3p_{3/2}$ 
levels are very close-lying.  
The calculated energy distance between the two levels is only 419 keV.   

The set of three levels, $1h_{11/2}$, $2f_{7/2}$ and $3p_{3/2}$, 
which strongly couple
each other by spin-independent quadrupole-quadrupole interaction, 
is the set of the $N$ = 5 oscillator major shell analogous to the set of 
three levels, $1g_{9/2}$, $2d_{5/2}$, and $3s_{1/2}$, of the $N$ = 4 major shell 
discussed in the previous subsection.  In the latter case a deformation 
may be energetically preferred 
for the system where some neutrons occupy the set of
levels due to the Jahn-Teller effect, while in the present case a deformation
may not be preferred since the one-particle levels for a moderate size of 
prolate shape in Fig. 4 
originating from $1h_{11/2}$ seem to mostly maintain the feature of the
single-j-shell.  
On the other hand, the occupation of the almost degenerate shells, $2f_{7/2}$ 
and $3p_{3/2}$, by some neutrons 
may lead to a deformed system for N = 83-90 when the
proton part of respective nuclei is energetically easily deformable 
and the pairing interaction plays a minor role. 
For example, the nucleus 
$^{127}_{44}$Ru$_{83}$ may be deformed, in a similar way to the nucleus 
$^{33}_{12}$Mg$_{21}$ in the ''island of inversion''(see Fig. 2.).  
If it is deformed the ground-state spin of the N = 83 nucleus is likely to be 
1/2$^-$ or 3/2$^-$ or 3/2$^+$ depending on $\beta$-values 
instead of 7/2$^-$ expected for spherical shape. 

In contrast to the three levels with parallel spin- and orbital-angular-momenta 
belonging to the $N$ = 5  oscillator major shell, filling   
the three levels, $1h_{9/2}$, $2f_{5/2}$ and $3p_{1/2}$, by some neutrons 
which belong to the same major
shell with anti-parallel spin- and orbital-angular-momenta may not lead to
deformation, in spite of the fact that 
those three levels may become almost degenerate around the Fermi level of 
certain nuclei as can be guessed from Fig. 4.

\section{CONCLUSIONS AND DISCUSSIONS}

A systematic study of the shell structure and the resulting possible
deformation around neutron-drip-line nuclei has been
carried out based on both bound
and resonant neutron one-particle energies obtained from phenomenological
Woods-Saxon potentials.  In order to solve the eigenvalue and eigenphase
problems for neutron one-particle bound and resonant levels, respectively, for a
given deformed potential, the
coupled differential equations obtained from the Schr\"{o}dinger equation are
integrated in coordinate space with correct asymptotic behavior.  
The coupling of a bound (or resonant) one-particle level with other levels, 
which are not obtained as resonant one-particle 
levels, is also
properly taken into account in the method of the present work. 

For spherical shape with the operator of the spin-orbit potential 
conventionally used, weakly-bound and/or low-lying resonant one-particle 
levels with parallel spin- and orbital-angular-momenta 
tend to gather together 
in the energetically lower half part of the oscillator major shell, 
while those levels with anti-parallel spin- and orbital-angular-momenta
gather in the upper half.  This grouping of energy levels 
in the spherical potential 
may lead to a possible deformation 
when neutrons start to occupy the lower half of
the major shell.   In contrast, the occupation of the upper half shell by
neutrons may not lead to a deformation.  

Some concrete result derived from the present study is: 
the magic numbers N=28 disappears and N=50 may disappear, while the magic
number N=82 may presumably survive. For spherical shape an appreciable amount
of energy gap appears at N=16 and 40.  Neutron-drip-line nuclei in the region of
neutron number above N=20, 40 and 82, namely N $\approx$ 21-28 (''island of
inversion''), N $\approx$ 41-54, and N $\approx$ 83-90, may be
quadrupole-deformed, though the possible deformation depends on the proton
number of respective nuclei.  

In actual nuclei it is possible that pair correlations may play an important
role in the determination of nuclear shape. 
It is generally understood that the pairing interaction tries to keep nuclei 
spherical, while the long-range part of two-body interactions such as
quadrupole-quadrupole interaction is responsible for deformation.  
In stable nuclei
deformed ground states are usually observed first after several nucleons filled 
one-particle levels above respective magic numbers.  
It is qualitatively understood that the presence of the several nucleons makes
the deformation induced by the long-range part of the interaction win against
the spherical shape preferred by the pairing interaction. 
However, it is noted that the first two low-lying $\ell_{j}$ shells above magic
numbers of stable nuclei do not couple strongly each other by the 
quadrupole-quadrupole interaction because there is a spin-flip between the two
$\ell_{j}$ shells.  For example, the $2p_{3/2}$-$1f_{5/2}$ shells just
above N = 28, the $2d_{5/2}$-$1g_{7/2}$ shells above N = 50 and the 
$2f_{7/2}$-$1h_{9/2}$ shells above N = 82. 
Therefore, the deformation-driving effect in stable nuclei 
by the presence of several nucleons above the magic numbers 
may be weaker than in the case of 
neutron-drip-line nuclei. In the latter case  
the deformation-driving force obtained by
filling neutrons in the almost degenerate $\ell_{j}$ shells that couple
strongly each other by quadrupole-quadrupole interaction  
may more easily win 
against the spherical shape preferred by the pairing interaction.  
On the other hand, the effect of pairing interaction which tries to keep nuclei
spherical can be also different in stable and neutron-drip-line nuclei, 
but the difference does not yet seem to be fully pinned down.  

In order to pin down a deformed shape of nuclei, 
the measurement of both the energy of the
lowest 2$^+$ state and $B(E2;2^+_1 \rightarrow 0^+)$ values in even-even nuclei
is important, but to observe the spin-parity of the ground state of odd-A nuclei
is often decisive.  For example, noting that the proton numbers Z = 21 (Sc), Z =
22 (Ti), Z = 23 (V) and Z = 24 (Cr) may help to have some deformation 
as is seen
from the fact that the energies of the two lowest-lying Nilsson one-proton
levels in the $pf$-shell decrease strongly as $\beta = 0 \rightarrow \beta > 0$,
the study of odd-N neutron-drip-line nuclei
with N = 41 such as $^{63}_{22}$Ti$_{41}$  
is highly desirable.  
In fact, the spin-parity 1/2$^-$ is already preliminarily assigned to the ground
state of $^{65}_{24}$Cr$_{41}$ \cite{BT10}.
On the other hand, the
experimental study of neutron-drip-line nuclei with N = 83 may not be possible
in a very near future.    
 
The possibility of deformation and the shell structure unique in 
neutron-drip-line nuclei, which are discussed in this article, 
should be duly studied 
by properly carrying out self-consistent Hartree-Fock (HF) calculations with 
appropriate effective interactions including pairing interaction. 
However, the effective interactions to be
used in HF calculations of neutron-drip-line nuclei are not yet properly fixed.
Moreover, such HF calculations have to be done by integrating the coupled
differential equations in 
coordinate space with proper asymptotic behavior of wave functions 
for $r = R_{max}$, at which both
nuclear potential and the coupling term are negligible, instead of using  
the expansion of wave functions 
in terms of harmonic oscillator bases or confining the system 
in a finite box.
This kind of proper HF calculations are not yet available for
neutron-drip-line nuclei. It remains
to be seen whether or not neutron-drip-line nuclei with certain neutron numbers
are actually deformed as suggested in the present work.

It is noted that the neutron one-particle 
states obtained from Nilsson diagrams for $\beta \neq$0 are 
those to be recognized as band-head configurations of odd-N nuclei.  
Thus, rotational states, which are
constructed based on those band-head states, should be in principle observed
using a proper experimental method, and those high-spin states will have 
narrow widths if they appear in the low-energy region.

A systematic change of the shell structure in the spherical potential discussed
in the present paper is strictly related to the characteristic feature of both 
the weakly-bound and 
low-lying resonant one-particle orbits with small $\ell$ values. 
The change of the shell structure and the resulting
one-particle energies in neutron-drip-line nuclei
must be taken into account in the shell-model
calculations, when shell model is applied to neutron-drip-line nuclei.

\newpage

\noindent
{\bf\large Figure captions}\\
\begin{description}
\item[{\rm Figure 1 :}]
Calculated neutron one-particle energies 
as a function of quadrupole deformation.
Parameters of the Woods-Saxon potential are chosen for the nucleus
$^{18}_{6}$C$_{12}$.  Bound one-particle energies at $\beta$ = 0 are $-$1.17 and
$-$0.66 MeV for the $1d_{5/2}$ and $2s_{1/2}$ levels, respectively, while
one-particle resonant $1d_{3/2}$ level is obtained at +4.36 MeV denoted by a 
filled circle.  One-particle
resonant levels for $\beta \neq$ 0 are not plotted unless they are important 
in the present discussion.  
For simplicity, calculated widths of one-particle resonant levels are not shown.
The neutron numbers, 8, 10 and 12, which
are obtained by filling all lower-lying levels, are indicated with circles.  
One-particle levels with $\Omega$ = 1/2, 3/2 and 5/2  are expressed by solid,
dotted and long-dashed curves, respectively, 
for both positive and negative parities.  The parity of levels can be seen from
the $\ell$-values denoted at $\beta$ = 0, $\pi$ = $(-1)^{\ell}$.
\end{description}

\begin{description}
\item[{\rm Figure 2 :}]
Calculated neutron one-particle energies 
as a function of quadrupole deformation.
Parameters of the Woods-Saxon potential are chosen for the nucleus 
$^{34}_{12}$Mg$_{22}$.  Bound one-particle energies at $\beta$ = 0 are $-$9.80,
$-$6.84, $-$4.75 and $-$0.24 MeV for the $1d_{5/2}$, $2s_{1/2}$, $1d_{3/2}$ and 
$1f_{7/2}$ levels, respectively, while one-particle resonant $2p_{3/2}$ and 
$1f_{5/2}$ levels are obtained at +0.15 and +6.18 MeV, respectively.  
The $2p_{1/2}$ one-particle resonant level is not obtained for the present
potential, however, its approximate position at $\beta$ = 0 is denoted by an
open circle, at which an eigenphase does not reach but comes close to $\pi$/2. 
One-particle resonant levels for $\beta \neq 0$ are not plotted unless they are
important in relation to the present interests.  
The neutron numbers, 20, 22 and 24, which are obtained by filling all
lower-lying levels, are indicated with circles.  
One-particle levels with $\Omega$ = 1/2, 3/2, 5/2 and 7/2 are expressed by
solid, dotted, long-dashed and dot-dashed curves, respectively, for both
positive and negative parities. 
\end{description}

\noindent
\begin{description}
\item[{\rm Figure 3 :}]
Calculated neutron one-particle energies as a function of quadrupole
deformation.
Parameters of the Woods-Saxon potential are chosen for the nucleus 
$^{66}_{22}$Ti$_{44}$.  Bound one-particle energies at $\beta$ = 0 are $-$8.82,
$-$5.54, $-$3.99, $-$3.94 and 
$-$0.48 MeV for the $1f_{7/2}$, $2p_{3/2}$, $2p_{1/2}$,
$1f_{5/2}$ and $1g_{9/2}$ levels, respectively, while one-particle resonant 
$2d_{5/2}$, $1g_{7/2}$ and $1h_{11/2}$ levels are obtained at +0.96, +5.66 and 
+7.57 MeV, respectively.  
The $2d_{3/2}$ one-particle resonant level is not obtained for the present
potential, however, its approximate position at $\beta$ = 0 is denoted by an
open circle, at which an eigenphase does not reach but comes close to $\pi$/2.  
The $3s_{1/2}$ resonant level does not exist in any case, but the open circle 
at $\beta = 0$ indicates the energy obtained by extrapolating 
the solid curve of the bound $\Omega^{\pi}$ = 1/2$^+$ orbit for $\beta > 0.12$
to $\beta$ = 0,  
though the calculated solid curve reaches zero 
at $\beta$ = 0.12 and cannot further continue to 
$\beta < 0.12$. 
The major component of the solid curve 
for $\epsilon_{\Omega} (< 0) \rightarrow 0$ is clearly $3s_{1/2}$. 
One-particle resonant levels for $\beta \neq 0$ are not plotted if they are not
relevant for the present discussion.  
The neutron numbers, 28, 40, 42, 44 and 48, which are obtained by filling all
lower-lying levels, are indicated with circles.  
One-particle levels with $\Omega$ = 1/2, 3/2, 5/2, 7/2 and 9/2 are expressed by
solid, dotted, long-dashed, dot-dashed and short-dashed curves, respectively,
for both positive and negative parities.  
\end{description}

\begin{description}
\item[{\rm Figure 4 :}]
Calculated neutron one-particle energies as a function of quadrupole
deformation.
Parameters of the Woods-Saxon potential are chosen for the nucleus 
$^{126}_{44}$Ru$_{82}$.  Bound one-particle energies at $\beta$ = 0 are $-$7.28,
$-$6.90, $-$5.70, $-$5.29, $-$4.05 $-$0.48 and 
$-$0.06 MeV for the $2d_{5/2}$, $1g_{7/2}$, $3s_{1/2}$,
$2d_{3/2}$, $1h_{11/2}$, $2f_{7/2}$ and $3p_{3/2}$ levels, respectively, 
while one-particle resonant 
$2f_{5/2}$, $1h_{9/2}$ $1i_{13/2}$ and $2g_{9/2}$ levels are obtained at +1.71, 
+1.91, +3.38 and +6.07 MeV, respectively.  
The $3p_{1/2}$ one-particle resonant level is not obtained for the present
potential, but the open circle 
at $\beta = 0$ indicates the energy obtained 
by calculating the spin-orbit splitting of the $\ell$ = 1 levels from that of
the $\ell$ = 3 ($2f_{5/2}$ and $2f_{7/2}$) levels and using the calculated 
energy of the $3p_{3/2}$ level.
One-particle resonant levels for $\beta \neq 0$ are not plotted if they are not
relevant for the present discussion.  
The neutron numbers 82 and 86, which are obtained by filling all
lower-lying levels, are indicated with circles.  
One-particle levels with $\Omega$ = 1/2, 3/2, 5/2, 7/2, 9/2 and 11/2 
are expressed by
solid, dotted, long-dashed, dot-dashed, short-dashed and dot-dot-dashed curves, 
respectively,
for both positive and negative parities.  
\end{description}

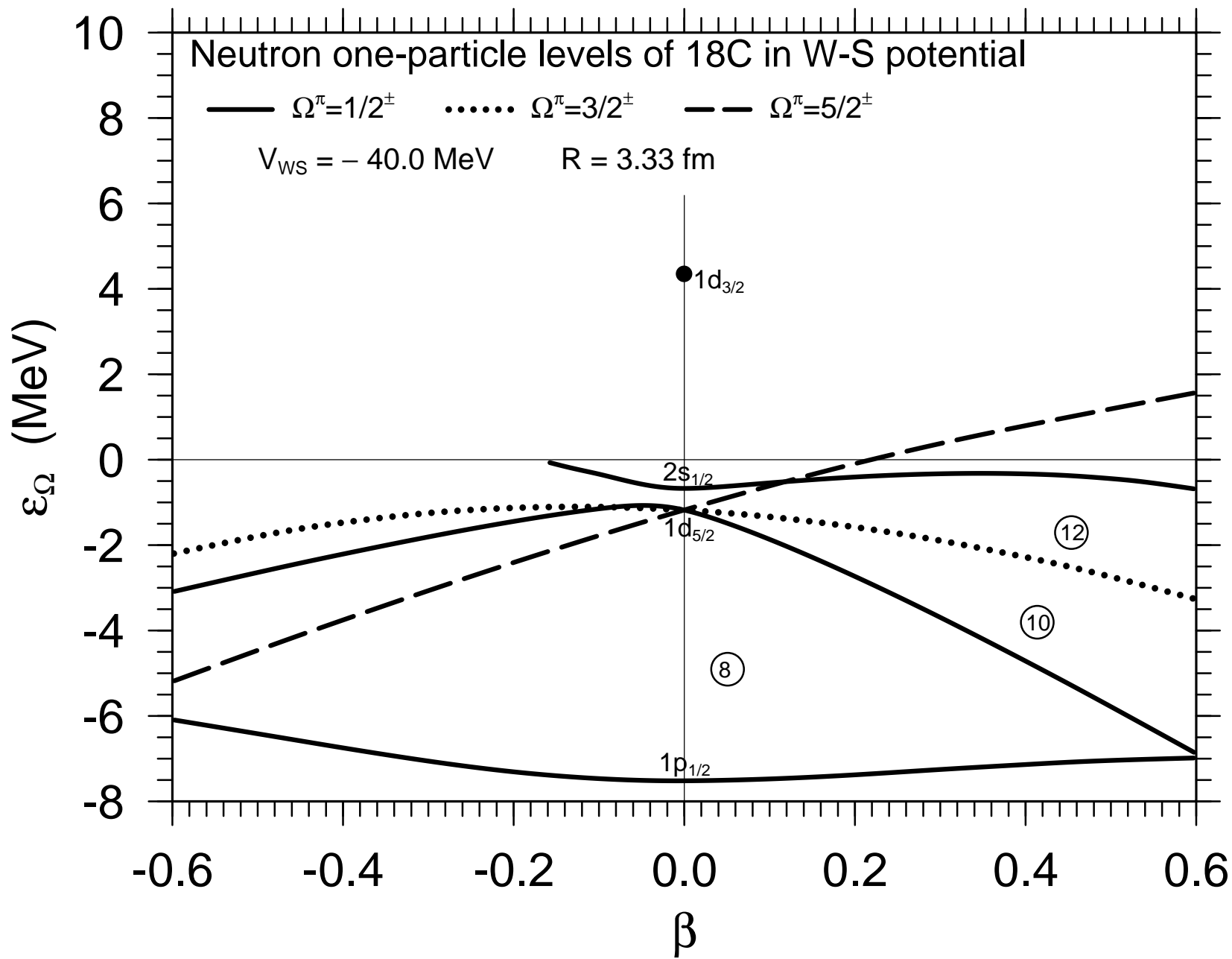
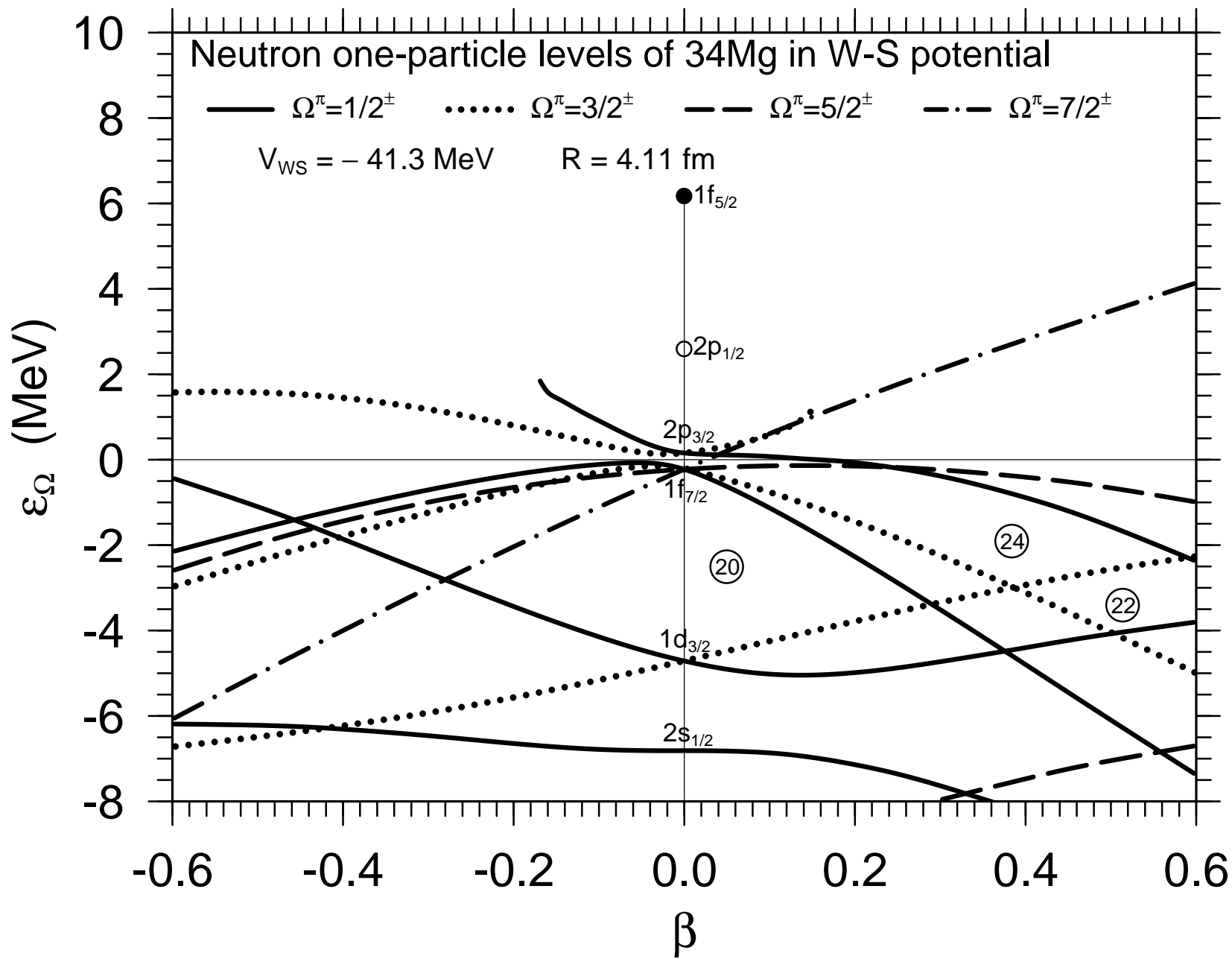
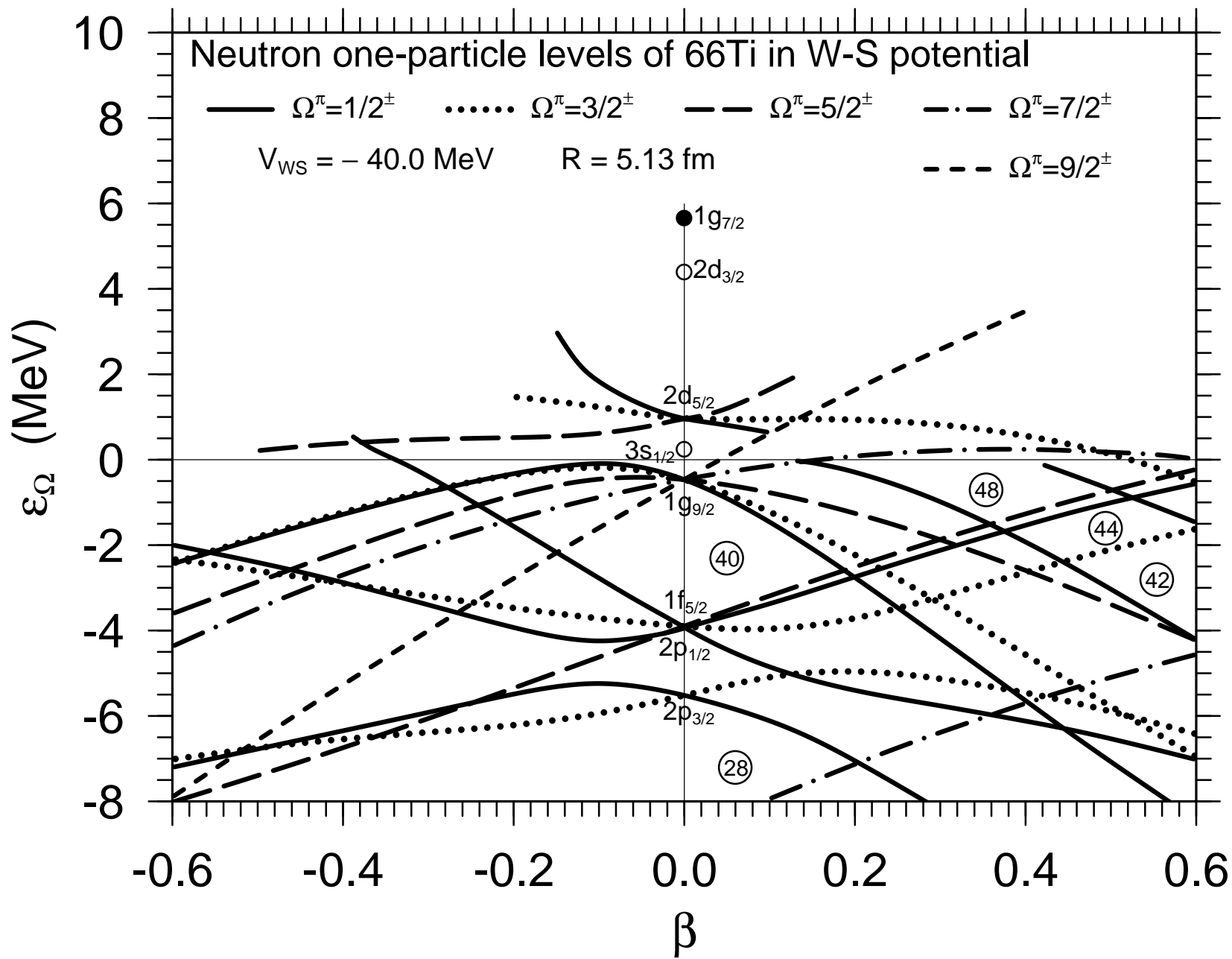
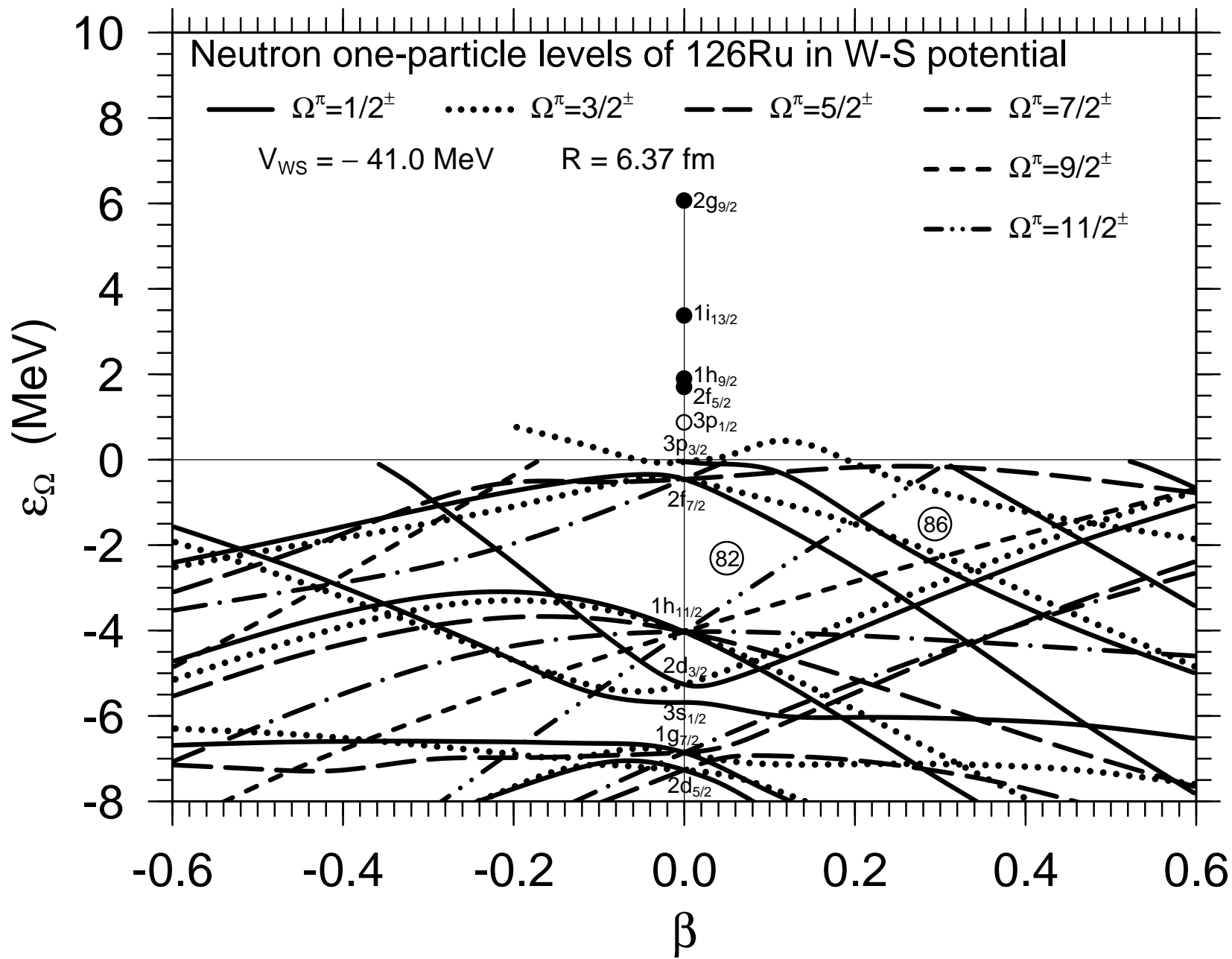
\end{document}